\documentclass[11pt,reqno]{article}
\usepackage[utf8]{inputenc}
\usepackage{amsmath, mathrsfs, amssymb, amsthm, amscd, amsfonts, graphicx, color,epsfig,latexsym,eucal,layout,fancyhdr, colonequals, enumerate, hyperref, fontenc, hyperref, graphics, a4wide, verbatim, psfrag}

\newcounter{EQNR}
\renewcommand{\contentsname}{Contents\\
{\footnotesize\normalfont(A table of contents should normally not be included)}}

\newtheorem{theorem}{Theorem}

\newtheorem{definition}[theorem]{Definition}

\newtheorem{lemma}[theorem]{Lemma}
\newtheorem{proposition}[theorem]{Proposition}

\newtheorem{remark}[theorem]{Remark}

\begin{document}

\title{An integer factorization algorithm which uses diffusion as a computational engine}
\author{Carlos A. Cadavid,\and Paulina Hoyos, \and Jay Jorgenson \and \and \and
Lejla Smajlovi\'{c}, \and Juan D. V\'elez}
\maketitle

\begin{abstract}\noindent
In this article we develop an algorithm which computes a divisor of an integer $N$,
which is assumed to be neither prime nor the power of a prime.  The algorithm
uses discrete time heat diffusion on a finite graph. If $N$ has $m$ distinct prime factors,
then the probability that our algorithm runs successfully is
at least $p(m) = 1-(m+1)/2^{m}$.   We compute the computational
complexity of the algorithm in terms of classical, or digital, steps and in terms of
diffusion steps, which is a concept that we define here.  As we will discuss below,
we assert that a diffusion
step can and should be considered as being comparable to a quantum step for an algorithm
which runs on a quantum computer.  With this, we prove that
our factorization algorithm uses at most $O((\log N)^{2})$ deterministic steps and at most
$O((\log N)^{2})$ diffusion steps with an implied constant which is effective. By comparison, Shor's algorithm is known to use at
most $O((\log N)^{2}\log (\log N) \log (\log \log N))$
quantum steps on a quantum computer.

As an example of our algorithm, we simulate
the diffusion computer algorithm on a desktop computer and obtain factorizations of $N=33$ and $N=1363$.

\end{abstract}

\section{Introduction}
Scientists continue to uncover ways in which nature manages and organizes itself.  For example,
the findings in the article \cite{Hig21} have been described as ``the first time we are seeing biology
actively exploiting quantum effects.''\footnote{\it Bacteria Know How to Exploit Quantum Mechanics to Steer Energy,
\rm SciTechDaily, March 29, 2021.} Whereas mathematics is sometimes
developed to study the physical world, there are times when the physical world provides a
guide for machine development.  In \cite{Cu17}, the authors cite
the article \cite{Ad94} as the foundational work on DNA computation and, furthermore,
assert that DNA computing can be viewed as a means to obtain the first physical design of a
non-deterministic universal Turing machine.  We refer the interested reader to the discussion in
section 3 of \cite{Cu17} where the authors given an intriguing, albeit brief, description of
molecular computing going back more than sixty years, beginning with \cite{Fey60}.

\vskip .10in
In this article we continue the theme of mathematically replicating the means by which
nature can be seen as a computational engine.  Specifically, we are interested in the manner
in which certain aspects of diffusion are simultaneously, not sequentially, observable.  For
instance, we offer the following self-evident observation.

\vskip .15in
\it
Assume that a heat source, such as a flame or a welding torch, is applied
to the center of a circular disc
of uniform thickness and material composition.  Then two observers who are measuring
temperatures at different points on the perimeter will detect a change of temperature at
their points of contact at the same rate.  \rm

\vskip .15in
The diffusion in the above setting is that of heat.  In this paper we study a
diffusion process which admits a different visualization which we will call a \it
diffusion ring, \rm or simply a \it ring. \rm

\vskip .15in
\it
Within a circular ring, imagine a beam of light $\mathcal{B}$ (or some type of focused energy)
emanating from a source at a perimeter point $\mathcal{P}_{0}$.  Upon contact with
another perimeter point $\mathcal{P}_{1}$ on the ring, the beam $\mathcal{B}$ splits into $M$ sub-beams
of equal magnitude in a prescribed set of directions toward perimeter points
$\mathcal{P}_{2,1}, \cdots \mathcal{P}_{2,M}$.  Let each sub-beam upon contact with
some $\mathcal{P}_{2,k}$ split in manner as similar to the reflection of $\mathcal{B}$ at $\mathcal{P}_{1}$,
and so on.  Then after $n$ such splittings, what portion of the original amount energy
has returned to $\mathcal{P}_{0}$?
\rm

\vskip .15in
The analysis in this article involves the mathematical understanding of the imagery of
a beam reflecting and splitting within a diffusion ring.  In this setting, we will count
a single \emph{diffusion step} as one instance of contact, reflecting
and subsequent splitting.  To be precise, the count will be mathematically captured
as one iteration of a symmetric
matrix on a finite dimensional vector space.
If each contact involves the splitting of a single beam into
$M$ sub-beams, then after $n$ diffusion steps, one will have $M^{n}$ paths of
light traversing the ring.   Though at this point we are solely interested in the mathematical
aspects of our set-up, one cannot help but imagine the  visualization of the diffusion ring.
Indeed, if such a ring were $1$ kilometer in diameter, and if the beam were to
travel at the speed of light, then after $0.01$ seconds one would expect to have
more than $M^{3000}$ sub-beams crossing various chords of the ring since
in almost all circumstances more than
$3000$ diffusion steps would have taken place.

\vskip .10in
Having established our definition of a diffusion step, we now can state the first main
result of this article.

\vskip .10in
\begin{theorem}\label{thm: main}
Let $N\geq 2$ be a positive integer which is assumed to be neither a prime nor a  prime
power.  Let $b$ denote an integer which is co-prime to $N$, and assume that the order of $b$
modulo $N$ is odd.  Then the order of $b$ modulo $N$ can be
computed in at most $O((\log N)^{2})$ diffusion steps.
\end{theorem}

\vskip .10in
As it turns out, there is an effective bound for the number of diffusion steps in Theorem \ref{thm: main}.
Indeed, we prove that
the number of diffusion steps $n$ which are needed in Theorem \ref{thm: main} satisfies the bound 
\begin{equation}\label{diffusion_bound}
n <  4\log N (\lfloor\log_2 N\rfloor+2),
\end{equation}
where $\lfloor \cdot \rfloor$ is the floor function, $\log$ is the natural logarithm function,
and $\log_{2}$ is the logarithm with base $2$.

\vskip .10in
Our second result stems from an application of Theorem \ref{thm: main} to the problem of
integer factorization.  Again, we assume that $N$ is neither a prime nor a prime power.
So, there is an integer $m \geq 2$ such that
$$
N = \prod_{i=1}^{m} p_i^{e_i}
$$
where the primes $p_{1},\cdots, p_{m}$ are distinct and the exponents $e_{1},\cdots, e_{k}$
are strictly positive.  Also, we will use the phrase \it digital steps \rm to signify the
term used to describe the computational complexity of an algorithm which is implemented on
a classical Turing machine.  With this, we can differentiate between \it digital steps \rm
and \it diffusion steps \rm which quantify the complexity of algorithms which
are carried out on either a digital computer or the above described diffusion machine.

\vskip .10in
\begin{theorem} \label{thm: integer fact}
Let $N$ be a positive integer with $m\geq 2$ distinct prime factors.
Then with probability $p(m) \geq 1 - (m+1)/2^m$ we can compute a non-trivial factor of $N$ in at most $O((\log N)^2))$ digital steps and at most
 $O((\log N)^{2})$ diffusion steps.
\end{theorem}

\vskip .10in
As with Theorem \ref{thm: main}, the bound \eqref{diffusion_bound} applies for the number of
diffusion steps in Theorem \ref{thm: integer fact}.  It is entirely possible that the number of
digital steps can be effectively bounded as well; however, for the sake of brevity of this article, we choose not
to study the effectiveness of the bound for the number of digital steps.

\vskip .10in
In case the algorithm terminates with no answer, one can simply repeat the computations.
Under the usual
assumptions of uniform
random selection, if we execute the algorithm $t$ times, then  the probability of
failing  is less than
$$
(1 - p(m))^t = \left((m+1)/2^{m}\right)^{t}.
$$
In other words, the probability of success becomes arbitrarily close to 1 with sufficiently
many implementations of the algorithm behind Theorem \ref{thm: integer fact}.

\vskip .10in
There exist deterministic algorithms which ascertain if $N$ is either a prime or the
power of a prime; see, for example, \cite{AKS04}, \cite{Be07}, or \cite{Ra80}.
In the two problems, the best known
algorithms have (classical) complexity of order $O((\log N)^{a})$ for some constant $a$.
With this, we do not view the assumption that $N$ is neither a prime nor a prime power as being restrictive, at least from the point of view of theoretical computability.

\vskip .10in

It is noteworthy that Shor's algorithm, which is the well-known
method for factoring using a quantum computer, takes at most
$O((\log N)^{2}\log (\log N) \log (\log \log N))$ quantum steps.  As with Shor's algorithm,
Theorem 2 has a certain probability which is less than one of a successful completion.  Initially, the probability of success for Shor's algorithm was determined to be at least $2/3$,
and more recent studies have sought to optimize the probability of success; see, for
example, \cite{Za13}.  Again, we will leave for elsewhere the problem of optimizing the
probability of success of Theorem \ref{thm: integer fact}.  In that regard, the methodology of
\cite{Za13} seems applicable.

\vskip .10in
There is current research into simulating Shor's algorithm on a digital computer;
see, for example, \cite{Mo16}, \cite{Pol09}, \cite{WHH17}.  In that vein,
we are able to readily simulate the diffusion computer
behind Theorem \ref{thm: integer fact}, and we provide two examples.  In the first, we
take $N=33$, and in the second we take $N=1363$.  The description of Theorem \ref{thm: integer fact} for these examples is given below, and the computer code which was written in
\it Maple \rm is provided in an Appendix to this paper.

\vskip .10in
Our approach to proving Theorem \ref{thm: main} is as follows.  Let $r$ denote the
order of $b$ modulo $N$, which we write as $r= \textrm{ord }_N b$.  Let $V$ denote
the set of powers of $b$ modulo $N$, so the cardinality of $V$ is $r$.  As the notation
suggests, the set $V$ is viewed as the set of vertices of a graph.  The edges of the graph
are formed by connecting each $b^{k}$ with points of the form $b^{k2^{j}}$ where $j$
ranges over positive and negative integers from $-(\lfloor \log_2 N\rfloor+1)$ to
$\lfloor \log_2 N\rfloor+1$.  We consider the diffusion process on the resulting graph
associated to the so-called half-lazy random walk; see section \ref{sec: random walks}
for details.  From \cite{Ba79} and \cite{Lo75}, we can express the
eigenvalues of the associated Laplacian in terms of
certain exponential sums.  As it turns out, optimal bounds for these exponential bounds
are known; see \cite{KM12} as well as \cite{Va09}.  When combining these results, we show that
after $O((\log N)^{2})$ diffusion steps one determines the cardinality of $V$, thus
the order of $b$ modulo $N$.

\vskip .10in
As we will discuss, the bounds we employ for such exponential sums are worst-case
scenarios.  As such, we expect that in practice fewer diffusion steps may suffice to
obtain some information about $r$.

\vskip .10in
Regarding Theorem \ref{thm: integer fact}, we follow the method which
is used in Shor's algorithm and replace the quantum computation step with
a diffusion computation.  In doing so, it is necessary to choose an integer $a$
whose order modulo $N$ is even.  For this, we prove how to reduce the problem, with
sufficiently high probability, to an implementation of Theorem \ref{thm: main}.

\vskip .10in
The outline of this paper is as follows.  In section 2 we establish notation
and recall necessary background material, including results from spectral graph theory
and discrete time heat kernels, which are used elsewhere.  Additionally, we formalize
the concept of \it diffusion process computing, \rm which stems from the ideas first
presented in \cite{HoRe20}.  In section 3 we obtain results using modular arithmetic
which are necessary in order to apply  Theorem \ref{thm: main} to prove Theorem \ref{thm: integer fact}.
The computations in section 3 appeal to known deterministic algorithms, such as the
Euclidean algorithm.  In section 4 we proof Theorem 1.  Specifically, we construct
the graph which is the mathematical realization of the above-described beam-splitting
reflection ring.  As stated, the bound on the number of diffusion steps in Theorem
\ref{thm: main} is obtained by certain eigenvalue bounds which in this case are
equivalent to bounds for exponential sums.  In section 5 we combine the results from
previous sections and complete the proof of Theorem \ref{thm: integer fact}.
In section 6 we describe the digital implementation of Theorem \ref{thm: integer fact}
and obtain the factorization of $N=33$ and $N=1363$, and in section 7 we present
a number of concluding remarks.

\vskip .10in
Finally, it is important to note that the ideas and methods of this paper were
motivated by the Master's Degree thesis \cite{HoRe20} written by one of the
authors (P. H.) of this article.  In \cite{HoRe20} the phrase \it heat computer \rm was coined,
and the idea formed our motivation for a \it diffusion computer. \rm  Additionally,
in \cite{HoRe20} it is shown how to construct heat computers which solve Simon's
problem and the Deutsch-Jozsa problem, and in each case the number of heat steps
coincides with the number of quantum steps for the known quantum algorithm solutions
of these problems.  In other words, the conceptualization of a diffusion computer began
with \cite{HoRe20}, and this article can be considered as a furtherance of the initial
ideas step forth in \cite{HoRe20}

\section{Preliminaries}

\subsection{Basic notation}
Any graph $X$ we consider in this article is finite, undirected and connected.
The set of vertices $V$ is finite, and the set of edges $E$ consists of a collection
of two-element subsets of $V$; we allow an edge to connect a vertex to itself, which
may be called a self-loop. We let $k=|V|$ denote the number of vertices of $X$.
Additionally, we assume $X$ has a real-valued weight function
$w \colon V\times V \to \mathbb{R}$ satisfying the following properties.
\begin{itemize}
    \item[(i)] {\bf Symmetry: }  For all $x, y \in V$, $w(x, y) = w(y, x)$.
    \item[(ii)] {\bf Semi-positivity:} For all $x, y \in V$, $w(x, y) \geq 0$.
    \item[(iii)] {\bf Positivity for Edges:} For all $x,y \in V$,
    $w(x, y) > 0$ if and only if $\{x,y\} \in E$.
\end{itemize}
The weight function generalizes the cases when there are multiple edges joining two vertices or self-loops.

\vskip .10in
Choose any ordering of the vertices.  The corresponding adjacency matrix $A$ of $X$ is a
$k \times k$ matrix whose
$(x,y)$-entry is given by the weight function, meaning that $A(x, y) = w(x, y)$.  From the above properties for the weight function, the adjacency matrix $A$ is symmetric with non-negative real entries.
The degree of a vertex $x \in V$ is defined as
$$
d(x) = \sum_{y \in V} w(x, y).
$$
A weighted graph $X$ is said to be regular of degree $d$ if $d(x) = d$ for all vertices $x \in V$.

\vskip .10in
\subsection{Random walks and the discrete time heat kernel} \label{sec: random walks}

We now assume that $X$ is a regular weighted graph of degree $d$.  A half-lazy random walk
on $X$ is a Markov chain  with state space $(V , \mathcal{P}(V))$ with arbitrary initial
 probability distribution $p_0 \colon V \to \mathbb{R}$, and transition probability matrix
 given by
\begin{align}\label{eq: W matrix}
     W \colonequals \frac{1}{2}\left(I + \frac{1}{d} A \right).
\end{align}
The matrix $W$ is called the half-lazy walk matrix of $X$. 
Intuitively, the half-lazy random walk is a process that starts with a single particle at some vertex, and at each step the particle either stays put at its current vertex with probability $1/2$ or moves randomly to a  neighbor with probability $1/2$.  In the second case, the particle moves from vertex $x$ to vertex $y$ with probability $w(x,y)/(2d)$.

\vskip .10in
Let $p_n \colon V \to \mathbb{R}$ denote the probability distribution at time $n$,
meaning after $n$ steps of the half-lazy random walk on $X$.  
Starting with an arbitrary probability distribution $p_0$ on $V$, then $p_{n}$ is
given inductively by
$$
 p_{n}(x) = \frac{1}{2}p_{n-1}(x) +
 \frac{1}{2}\sum_{y \in V} \frac{w(x,y)}{d} p_{n-1}(y).
$$
Equivalently, we can view $p_n$ as a column vector from $\mathbb{R}^k$,
 so this equation can be
written in matrix form as
\begin{align*}
    p_{n} = \frac{1}{2}\left( I+\frac{1}{d}A \right) p_{n-1} = W p_{n-1} = W^n p_0.
\end{align*}

Let us denote the $(x,y)$-entry of $W^{n}$ by $w_{n}(x,y)$, which can be interpreted as the probability that a particle which follows the half-lazy random walk on $X$ and
starts at $y$ is at vertex $x$ after $n$ steps.

\vskip .10in
The function $w_{n}(x,y)$
is called the discrete time heat kernel of $X$  because the random walk provides a
probabilistic interpretation of heat diffusion in $X$, where the temperature at a vertex
 $x$ is considered to be a manifestation of ``heat particles'' which spread randomly in all
 directions. As such, one can view $p_n$ as the distribution of these heat particles at time $n$.
 That is, if one starts with $m_y$ units of heat at each vertex $y \in V$,
 then the temperature at vertex $x$ after $n$ steps is $p_n(x)= \sum_{y\in V}m_y w_n(x,y)$.



\vskip .10in
Under this interpretation, the physical principle of conservation of energy can be stated as
\begin{align*}\label{eq: difference equation}
p_{n+1}-p _{n}=(W-I)p_{n}.
\end{align*}
The difference $\partial_{n}p_{n} \colonequals p_{n+1}-p_{n}$ is called the discrete time derivative, and the operator $\Delta \colonequals W-I$ is referred to as the discrete  Laplacian.  With this notation the above equation reads
\begin{equation} \label{eq: discrete heat eq}
\partial_{n}p_{n}=\Delta p_n,
\end{equation}
which is known as the discrete heat equation on $X$.

\vskip .10in
The standard solution of the discrete heat equation with
initial condition $p_{0}$ is $p_{n}=W^n p_0$, and the solution can be expressed by
diagonalizing the symmetric matrix $W$.  Specifically, let $\lambda _{0},\dots ,\lambda _{k-1}$ be the 
eigenvalues of $W$ with corresponding eigenvectors $\psi_{0},\dots ,\psi_{k-1}$ which
form an orthonormal basis for $\mathbb{R}^k$.  Then, using elementary linear algebra, one
obtains that

\begin{equation*}
p_n=\sum_{j=0}^{k-1}\langle \psi_j, p_n \rangle \psi _{j}=\sum_{j=0}^{k-1}\langle \psi_j, W^n p_0 \rangle \psi _{j}=\sum_{j=0}^{k-1}\langle \psi_j, p_0 \rangle \lambda _{j}^{n}\psi _{j},
\end{equation*}
where $\langle \cdot, \cdot \rangle$ denotes the standard scalar product of vectors in
$\mathbb{R}^k$, meaning that if $\psi_j(x)$ denotes the $x$-entry of $\psi_j$, then
$$
\langle \psi_j, p_0\rangle = \sum_{x\in V} \psi_j(x) p_0(x).
$$

\vskip .10in
In our notation, $\lambda_{0} = 1$ and $\vert \lambda_{j} \vert < 1$ for all $j = 1, \cdots, k-1$.  Also, $\psi_{0}(x) = 1/k$ for all $x \in V$.
As a result, if $n$ goes to infinity, the solution $p_n$ converges to
the uniform probability distribution on the set $V$ regardless of the initial condition $p_0$.  More precisely, we have the following proposition.

\vskip .10in
\begin{proposition} \label{prop: bound for the heat flow}
With the notation as above, let $\lambda_1$ be the largest eigenvalue of $W$ less than 1.
Assume that the initial condition $p_{0}$ is a probability distribution, meaning it
is semi-positive and has $\ell^{1}$ norm equal to one.  Then for all $x\in V$ and all
$n\geq 0$ we have that
\begin{align*}
    \left\vert p_n(x) - \frac{1}{k} \right\vert \leq \lambda_1^n.
\end{align*}
\end{proposition}


\vskip .10in
\begin{proof}
From \eqref{eq: W matrix} we have that a vector $\psi$ is an eigenvector of $A$ with eigenvalue $\eta$ if and only if $\psi$ is an eigenvector of $W$ with eigenvalue
\[ \lambda = \frac{1}{2}\left(1 + \frac{1}{d} \eta \right). \]
The eigenvalues of the adjacency matrix $A$ of a degree $d$ regular weighted graph lie in the interval $[-d,d]$ with $d$ being the largest eigenvalue;
see for example Theorem 7.5 of \cite{Ni18}. Hence the eigenvalues of $W$ satisfy the inequalities
$$
1 = \lambda_0 > \lambda_1 \geq \cdots \lambda_{k-1} \geq 0.
$$
The eigenvector $\psi_{0}$ corresponding to $\lambda_{0}$ is such that
$\psi_0(x) = 1/k$ for all $x \in V$.
With this, we have for any $x \in V$ and all $n \geq 0$ the expansion 
\begin{align} \label{eq: p_n computation}
    p_n(x) &= W^n p_0(x) = \langle \psi_0, p_0 \rangle \psi_0(x) + \sum_{j=1}^{k-1}\langle \psi_j, p_0 \rangle \lambda _{j}^{n}\psi_{j}(x) = \frac{1}{k} + \sum_{j=1}^{k-1}\langle \psi_j, p_0 \rangle \lambda _{j}^{n}\psi_{j}(x),
\end{align}
where the last equality follows from the assumption that $p_{0}$ is a probability
distribution.

\vskip .10in
Since the set of eigenvalues are orthonormal, we have a version of Parseval's formula, namely 
\begin{align} \label{eq: l2 bound for distance}
    \left\vert p_n(x) - \frac{1}{k} \right\vert^2 
    \leq
    \sum_{y \in V} \left\vert p_n(y) - \frac{1}{k} \right\vert^2 
    = \sum_{j=1}^{k-1} \left( \langle \psi_j, p_0 \rangle \lambda _{j}^{n} \right)^2 \leq \lambda_1^{2n} \sum_{j=1}^{k-1}  \langle \psi_j, p_0 \rangle^2.
\end{align}
Moreover, by writing $p_0=\sum\limits_{j=0}^{k-1} \langle \psi_j, p_0 \rangle \psi_j$
we obtain
$$
\sum_{j=1}^{k-1}  \langle \psi_j, p_0 \rangle^2\leq \sum_{j=0}^{k-1}  \langle \psi_j, p_0 \rangle^2 =\langle p_0, p_0 \rangle =\sum_{x\in V} p_0(x)^2 \leq \sum_{x\in V} p_0(x)=1.
$$
The last inequality follows from the fact that $p_0$ is a probability distribution on $V$. Combining this inequality with \eqref{eq: l2 bound for distance} we get
$$
  \left\vert p_n(x) - \frac{1}{k} \right\vert^2 \leq \lambda_1^{2n},
$$
which completes the proof of the assertion.
\end{proof}

\vskip .10in
\subsection{Weighted Cayley graphs of finite abelian groups}
Let $G$ be a finite abelian group, and let $S\subseteq G$ be a fixed symmetric subset
generating $G$. The symmetry condition means that if $s\in S$ then $-s\in S$. Moreover, let
$\alpha \colon S \to \mathbb{R}^{>0}$ be a function such that $\alpha(s) = \alpha(-s)$.
One can construct a weighted Cayley graph $X= \text{Cay}(G, S, \alpha)$ of $G$ with respect
to $S$ and $\alpha$ as follows.
The vertices of $X$ are the elements of $G$. Two vertices $x$ and $y$  are connected with an edge if and only if $x-y \in S$.  The weight $w(x,y)$ of the edge $(x,y)$ is
$w(x,y) = \alpha(x-y)$.  With all this, one can show that $X$ is a
regular weighted graph of degree
$$
 d= \sum_{s \in S} \alpha (s).
 $$
By assuming that $S$ generates $G$ it follows that the Cayley graph $X$ is connected.

\vskip .10in
Given $x \in G$, let $\chi_x$ denote the character of $G$ corresponding to $x$;
see, for example, \cite{CR62}.  Then $\chi_x$ can be represented as an
eigenvector of the adjacency operator $A$ of $X$ with corresponding eigenvalue equal to
\begin{equation*}
\eta_{x}=\sum_{s\in S} \alpha(s) \chi_{x}(s);
\end{equation*}
see Corollary 3.2 of \cite{Ba79}. It follows that  $\chi_{x}$ is an eigenvector of the half-lazy walk matrix $W$ of $X$ with eigenvalue
\begin{equation*}
\lambda_{x}=\frac{1}{2}\left(1+\frac{1}{d}\eta_{x}\right).
\end{equation*}%

\vskip .10in
Let us number the \emph{distinct} eigenvalues of $W$ as
\begin{equation*}
1=\lambda _{0}>\lambda _{1}>\dots >\lambda _{l}\geq 0,
\end{equation*}
where, obviously $1\leq l\leq |G|-1$.

\vskip .10in
The results of section \ref{sec: random walks} on half-lazy random walks on the graph $X$ apply to deduce that for any initial probability distribution $p_0$ on the set $G$ of vertices of $X$, the half-lazy random walk $\{p_n\}_{n=0}^\infty$ converges to the uniform distribution on $G$ as $n\to\infty$. More precisely, for all $x \in G$ and all positive integers $n$, the inequality
\[ \left\vert p_n(x) - \frac{1}{\vert G \vert} \right\vert \leq \lambda_1^n \]
holds true.

\vskip .10in
Moreover, fix $1\leq i\leq l$, and let $E_{i}\subseteq \mathbb{R}^{|G|}$ be the eigenspace corresponding to the eigenvalue $\lambda _{i}$. The set $B_{i}=\{\chi_x:\lambda _{x}=\lambda _{i}\}$ is an orthonormal basis for $E_{i}$, so the projection $h_{i}$ of $p_0$ onto the eigenspace $E_{i}$ is given by
\begin{equation*}
h_{i}=\sum_{\substack{ x\in G  \\ \lambda _{x}=\lambda _{i}}}\left\langle
\chi_{x},p _{0}\right\rangle_G \chi_{x},
\end{equation*}
where $\langle \cdot, \cdot \rangle_G$ denotes inner product on the set all complex-valued functions on $G$ and is defined as
\[ \langle f,g \rangle_G \colonequals \frac{1}{\vert G \vert }\sum_{x \in G} \overline{f(x)}g(x).\]
With this in mind, the solution $p_{n}$ of the discrete heat equation \eqref{eq: discrete heat eq} on $X$ subject to the initial probability distribution $p _{0}\colon G\rightarrow \mathbb{R}$ can also be written as
\begin{equation}
p_{n}=\sum_{i=0}^{l}\lambda _{i}^{n}\text{ }h_{i}.
\end{equation}

\vskip .10in
Expressing the solution in terms of the projections of the initial condition onto the
eigenspaces of $W$ has the advantage that it is not subject to a particular basis choice.
 As we shall see, this is particularly useful for defining the notion of a \it Diffusion
  Computer. \rm

\vskip .10in
\subsection{Diffusion process computing}
A diffusion process in $X=\text{Cay}(G,S, \alpha)$ may be regarded as an analog computation on $X$ in the following sense.

\vskip .10in
\begin{definition} \label{def: diffusion computing}
A real-valued function $h$ on $G$ is said to be \textrm{computable by a diffusion process in} $X$ with initial condition $p_0 \colon G \to \mathbb{R}$ if the following holds.
Let $\{p_{m}\}_{m=0}^\infty$ be the solution to the discrete heat
equation \eqref{eq: discrete heat eq} in $X$ with initial probability distribution $p_0$.
Then for any given $\varepsilon >0$ there exists a positive integer $n = n(\varepsilon)$
such that for all $m > n(\epsilon)$ and all $x \in G$, we have that
$$
\left\vert p_{m}(x)-h(x)  \right\vert < \varepsilon.
$$
\end{definition}

\vskip .10in
Colloquially, we will refer to $X$ as a \it Heat Computer \rm or descriptively as a
\it Diffusion Computer. \rm The function $h$ will be called \it diffusion computable. \rm
As stated, the concept of a Diffusion Computer first was
developed in \cite{HoRe20}.  By the linearity of the discrete heat equation and the uniqueness of its solution, any linear combination of diffusion-computable functions on $X$ is also computable by a diffusion process in $X$.

\vskip .10in
\begin{theorem}
\label{heat computable}Let $f$ be an arbitrary real-valued function on $G$, and let $h_{0},h_{1}\dots ,h_{l}$ be the corresponding projections of $f$ onto the eigenspaces $E_{0},E_{1}\dots ,E_{l}$ of the half-lazy walk matrix $W$. Then the functions $h_{0},h_{1}\dots ,h_{l}$ are diffusion computable on $X=\mathrm{Cay}(G,S, \alpha).$
\end{theorem}

\vskip .10in
A formal proof is in \cite{HoRe20}. The computation is carried out by induction and can be described in terms of the heat diffusion. First one computes $h_{0}$ by just letting heat
 flow for enough time and then reading the temperature of the steady-state solution. We
  repeat the same procedure, this time with initial temperature function $\lambda
  _{1}^{-n_\ast}(f-h_{0})$ for a suitable value of $n_\ast$. The fact that the eigenvalues
   are strictly decreasing guarantees that the steady-state solution converges to $h_{1}$, and we can continue recursively in this manner.

   It is important to note that in the Definition \ref{def: diffusion computing} above and throughout this article we use the expression ``diffusion process" instead of the ``heat computer" which was coined in \cite{HoRe20}.  We do so because the principle of computation stays the same if the heat is replaced by any other diffusion process.

\vskip .15in
\it
Any physical device capable of implementing the recursive procedure described in Theorem \ref{heat computable} will be called a diffusion based computer, or simply a diffusion computer.  \rm

\vskip .15in
In essence, a diffusion  computer is a device that computes the Fourier expansion of a complex-valued function $f$ defined on $G$ in its base of characters. 
The Fourier spectrum of $f$ can be obtained as follows. Fix $1\leq i\leq l$, and suppose that the eigenspace $E_{i}$ has dimension $m_{i}$. Recall that
\begin{equation*}
h_{i}(x)=\sum_{\substack{ a\in G  \\ \lambda _{a}=\lambda _{i}}}\left\langle
\chi_{a},f\right\rangle_G \chi_{a}(x), \quad x\in G.
\end{equation*}
Therefore, by first computing $h_{i}$ and then evaluating it at $m_{i}$ different elements of $G$ we get a system of linear equations that allows
us to solve for the Fourier coefficients $\left\langle \chi_{a},f\right\rangle_G $ such that $\lambda _{a}=\lambda _{i}$. By letting $i$ vary, we
get the whole spectrum of $f$.

\vskip .10in
It is an extraordinary fact that each one of the iterations is carried out by an extremely simple repetitive procedure, namely that the value of the function
at each vertex is replaced by the average value of its neighbors, and this same thing occurs at every vertex. In many cases the first projections
are enough to infer interesting properties about $f$. For that matter, in this article we only exploit the capability of computing the zeroth projection
of $f.$

\vskip .10in

\subsection{Equating quantum steps and diffusion steps}

In \cite{Ma21}, the author quotes the principal manager of the quantum-computing group at Microsoft Research in Redmond, Washington as saying that
``Quantum computing is essentially matrix vector multiplication — it’s linear algebra underneath the hood''.  From the linear algebra point of view, it is the opinion of  the authors of this article that, in general, a diffusion computer can be regarded as the $\ell^{1}$ version of the
quantum computer, which itself is based on $\ell^{2}$ theory. Let us describe the meaning of this comment.

\vskip .10in
As it is known,
a quantum computation entails two different types of operations. The first is a deterministic operation which is just the abstract version of the classical evolution equation in quantum mechanics.  For this, a unitary vector $\phi$ in a Hilbert space $(\mathbb{C}^{n},\ell^{2})$
evolves into a new vector $\psi = U \phi$ where $U$ is some unitary operator.  The second operation involves a measurement of this new state. This procedure is non-deterministic and has the effect of ``collapsing $\psi$''.  More precisely, each particular measurement is
modeled by the decomposition of $\mathbb{C}^{n}$ into finite orthogonal subspaces $H_{i}$. By ``collapsing'' we mean composing $\psi$ with the projection $\psi_{i}$ onto $H_{i}$, and this
projection occurs with probability $\vert \psi_{i}\vert^{2}$.

\vskip .10in
By comparison, on a diffusion computer the unitary vector $\phi$ is replaced by a stochastic
vector (i.e. $\ell^1-$norm one vector) in $(\mathbb{R}^{N},\ell^{1})$. The evolution of $\phi$ is determined by
the symmetric operator $W$.  Then it is reasonable to define one diffusion computation step as one application of $W$
which maps $\phi$ to $W\phi$.  A measurement, on the other hand, is just a classical inspection
of the vertices of the graph $X$ where the diffusion process takes place.  It is evident
that the inspection of a subset of elements of $X$ is also a projection operation.  As with
a quantum computer, a projection also is counted as one step.

\vskip .10in
\it In summary, one determines the steps in a quantum computation by counting the number of
compositions of unitary matrices and projections, and in a diffusion process by counting the number of
compositions of symmetric matrices and projections.  As such, we consider a count of
quantum steps to be comparable to a count of diffusion steps. \rm

\section{Some number theoretic considerations} \label{sec:elementary NT}

As before, we let $N$ be a positive, odd integer which we write as a product
\begin{equation}\label{eq: N=product of primes}
N = \prod_{i=1}^{m} p_i^{e_i}
\end{equation}
with $m\geq 2$ different odd prime factors with exponents $e_i > 0$.  As such, we assume
that $N$ is not prime and not a prime power.
Let $\mathbb{Z}_{N}$ denote the set of inequivalent classes of
integers modulo $N$.  There is a natural mapping
\begin{equation}\label{CRT}
\mathbb{Z}_{N} \to \mathbb{Z}_{p_{1}^{e_1}}\times  \cdots \times \mathbb{Z}_{p_{m}^{e_m}}
\end{equation}
defined by
$$
a \mapsto (a \textrm{\rm \,mod\,} p_{1}^{e_1}, \cdots, a \textrm{\rm \,mod\,} p_{m}^{e_m}).
$$
By the classical Chinese Remainder Theorem, \eqref{CRT} is an isomorphism. Furthermore, the mapping \eqref{CRT} yields an isomorphism of the respective commutative rings, which we write as
\begin{equation}\label{gN}
g_N \colon \mathbb{Z}_{N}^{\ast }\rightarrow \mathbb{Z}_{p_{1}^{e_1}}^{\ast }\times \cdots \times \mathbb{Z}_{p_{m}^{e_m}}^{\ast}.
\end{equation}
In a slight abuse of notation, we occasionally use $x$ to denote either an element in
$\mathbb{Z}_{N}^{\ast }$ or its image $g_{N}(x)$.
For each $i$, the ring $\mathbb{Z}_{p_{i}^{e_i}}^{\ast }$ is a cyclic group under multiplication.
Let us denote its generator by $u_i$.  We shall write the order of $u_{i}$ as
$$
\mathrm{ord}_{p_{i}^{e_i}}u_i=p_i^{e_i-1}(p_i - 1)= 2^{c_i}p_i^\prime
$$
where $c_i > 0$ and $p_i^\prime$ is an odd positive integer.

\vskip .10in
{\it Without loss of generality we assume that the primes $p_1,\ldots,p_m$ are
ordered so that we have the inequalities $c_1\geq c_2 \geq \dots \geq c_m$.}

\vskip .10in
The following proposition computes, under certain circumstances, a non-trivial
square root of 1 modulo $N$.

\vskip .10in
\begin{proposition} \label{prop: main elementary NT}
For any  $a\in  \mathbb{Z}_{N}^{\ast }$, let  $s \geq 0$ be the \textrm{least power}  of $2$ such that
$$
a^{2^{s}q} \equiv 1 \textrm{\rm \,mod\,} N
$$
for some odd integer $q$. Let us write
$$
g_N(a)=(u_{1}^{d_{1}},\dots, u_{m}^{d_{m}}).
$$
If there exist $i,j$ with $1\leq i<j \leq m$ such that $d_{i}$ is odd and $d_{j}$ is even, then $s>0$.
Furthermore, if we set
$x =a^{2^{s-1}q} $
then
$$
x^{2} \equiv 1 \textrm{\rm \,mod\,} N
\,\,\,\,\,\,\,\,
\textrm{\it and}
\,\,\,\,\,\,\,\,
 x\not\equiv \pm 1 \textrm{\rm \,mod\,} N.
$$
\end{proposition}

\vskip .10in
\begin{proof}
By assuming $d_{i}$ is odd and $d_{j}$ is even, we can write $d_{j}=2^{v}d_{j}^{\prime }$
where $v>0$ and $d_{j}^{\prime }$ is an odd integer.

\vskip .10in
Because $u_{i}^{2^{s}qd_{i}}\equiv 1$ in $\mathbb{Z}_{p_{i}^{e_i}}^{\ast }$ the order of $u_{i},$ which is $2^{c_{i}}p_{i}^{\prime },$ divides $2^{s}qd_i$. This implies that $s\geq c_{i} > 0$ and $p_{i}^{\prime }$ divides $qd_{i}.$  Thus $s\geq 1$.  Set
$$
x=a^{2^{s-1}q}=(u_{1}^{2^{s-1}q d_1}, \dots , u_{m}^{2^{s-1}q d_m}).
$$
Then
$$
x^{2}=(u_{1}^{2^{s}qd_{1}}, \dots, u_{m}^{2^{s}qd_{m}}) \equiv 1 \textrm{\rm \,mod\,} N
$$
Moreover, by the minimality of $s$, we have that $x\not\equiv 1 \textrm{\rm \,mod\,} N$.
It remains to show that $x\not\equiv -1 \textrm{\rm \,mod\,} N$.

\vskip .10in
If $x \equiv -1 \textrm{\rm \,mod\,} N$, then for every $k$ we have that
$x \equiv -1 \textrm{\rm \,mod\,} p_{k}^{e_{k}}$.  Therefore, if
$x \not\equiv -1 \textrm{\rm \,mod\,} p_{k}^{e_{k}}$ for a single index $k$, then
$x \not\equiv -1 \textrm{\rm \,mod\,} N$.  Indeed, we now will prove that
$x \not\equiv -1 \textrm{\rm \,mod\,} p_{j}^{e_{j}}$ for the particular index $j$ for which $d_{j}$
is odd and $j > i$, which is assumed to exist as stated in the premise of the proposition.
More specifically, we claim that $x \equiv 1 \textrm{\rm \,mod\,} p_{j}^{e_{j}}$, which we
rewrite as
\begin{equation}\label{e2}
u_{j}^{2^{s-1}qd_{j}} \equiv u_{j}^{2^{s+v-1}qd_{j}^{\prime }} \equiv 1 \, \textrm{\rm \,mod\,} p_{j}^{e_{j}}.
\end{equation}

\vskip .10in
Equation \eqref{e2} holds if and only if $2^{c_j}p_{j}^{\prime }$,
the order of $u_j$ divides $ 2^{s+v-1}qd_{j}^{\prime }$. Equivalently,
equation \eqref{e2} holds
if and only if
\begin{equation}\label{div_cond}
s+v-1\geq c_j
\,\,\,\,\,\,\,\,
\textrm{\rm and}
\,\,\,\,\,\,\,\,
p_{j}^{\prime } \mid qd_{j}^{\prime }.
\end{equation}
Because $u_{j}^{2^{s}qd_{j}}\equiv 1 \, \textrm{\rm \,mod\,} p_{j}^{e_{j}}$,
the order of $u_{j}$, which is $2^{c_j}p_{j}^{\prime },$ divides
$2^{s}qd_{j}=2^{s+v}qd_{j}^{\prime }$.  Since all integers $p_{j}', q$ and $d_{j}'$ are odd,
we then have that $s+v\geq c_j$ and $p_{j}^{\prime }
 \mid qd_{j}^{\prime }$.  Thus, the second condition in \eqref{div_cond} is proved.
Since $s\geq c_{i}\geq c_{j}$ and $v>0$, we have that
$$
s+v-1\geq c_{i}+v-1\geq c_j,
$$
which proves the first condition in \eqref{div_cond} and completes the proof of the proposition.
\end{proof}

\vskip .10in
\begin{lemma} \label{lem: repetitions in S}
Let $N$ be a product of $m\geq 2$ distinct odd primes,  and let $M=\lfloor \log _{2}N \rfloor+1$.
For any $a \in \mathbb{Z}_N^{\ast}$, define the subset $S=S(a)$ of $\mathbb{Z}_N^{\ast}$ by
$$
S=\{a^{ \pm 2^{t}} \textrm{\rm \, mod\, } N: \textrm{\rm \,for all \,}t=0,\ldots ,M\}.
$$
If there are repetitions in $S$, then with probability $p(m) = 1 - (m+1)/2^m$ we can find
an $x \in \mathbb{Z}_N^{\ast}$ in at most $O(\log_2 N)$ deterministic steps
for which
\begin{equation}\label{x_condition}
x^{2} \equiv 1 \textrm{\rm \,mod\,} N
\,\,\,\,\,\,\,\,
\textrm{\it and}
\,\,\,\,\,\,\,\,
 x\not\equiv \pm 1 \textrm{\rm \,mod\,} N.
\end{equation}
\end{lemma}

\vskip .10in
\begin{proof}
Let us assume there is at least one repetition in $S$, meaning that for
some $\ell$ and $\ell'$ we have that $a^{2^{l}}\equiv a^{\pm 2^{l^{\prime }}}\textrm{\rm \,mod\,} N $.
Without loss of generality, we may assume that $l>l^{\prime }$. In other words,
we have that
\begin{equation}\label{repetition}
a^{2^{l} \pm 2^{l^{\prime }}}\equiv 1 \textrm{\rm \,mod\,} N.
\end{equation}
Solving this equation, we obtain
$$
a^{2^{l^{\prime }}(2^{l-l^{\prime }} \pm 1)}\equiv 1 \textrm{\rm \,mod\,} N
\,\,\,\,\,\,\,
\textrm{\rm so then}
\,\,\,\,\,\,\,
a^{2^{l^{\prime }}q}\equiv 1 \textrm{\rm \,mod\,} N,
$$
where $q=2^{l-l^{\prime }} \pm 1$, which is an odd integer.

\vskip .10in
Once the set $S$ has been constructed, and once one is given the values of $\ell$ and $\ell'$
in \eqref{repetition}, then
one can determine the smallest $s$ such that $a^{2^{s}q}\equiv 1 \textrm{\rm \,mod\,} N$
in at most $O(\log_2 N)$ deterministic computations.

\vskip .10in
As stated, the
mapping \eqref{gN} is a bijection between $\mathbb{Z}_N^\ast$ and $\mathbb{Z}_{p_{1}^{e_1}}^{\ast }\times \cdots \times \mathbb{Z}_{p_{m}^{e_m}}^{\ast}$.  Let
$\{u_1,\ldots,u_m\}$ denote a set of generators
of the cyclic multiplicative groups $\mathbb{Z}_{p_{1}^{e_1}}^{\ast }, \ldots, \mathbb{Z}_{p_{m}^{e_m}}^{\ast}$.  With this, we conclude that
any element $a\in \mathbb{Z}_N^\ast$ is uniquely determined by the set of exponents $(d_1,\ldots,d_m)$ for which $g_N(a)\equiv (u_{1}^{d_{1}},\dots, u_{m}^{d_{m}})$.
In other words, choosing a random element $a\in\mathbb{Z}_N^\ast$ is equivalent to choosing a random $m$-tuple $(d_1,\ldots,d_m)$, where each $d_i$ is randomly and uniformly chosen from the set $\{1,\ldots, p_i^{e_i-1}(p_i-1)\}$, $i=1,\ldots,m$.  Note that the set of
integers from $1$ to  $p_i^{e_i-1}(p_i-1)$ has equal number of even and odd elements.

\vskip .10in
With the above discussion, we can now estimate the probability $p(m)$.  Proposition \ref{prop: main elementary NT} proves that when $s>0$ and if we set $x=a^{2^{s-1}q}$,
then \eqref{x_condition} holds when $d_{i}$ is odd and $d_{j}$ is even for some $1\leq i<j \leq m$.  Assuming $a$ is chosen randomly and uniformly, let
us show that the probability that $a$ satisfies the conditions on $i$ and $j$, hence $d_{i}$ and $d_{j}$, by showing that the probability of
the complementary event is $(m+1)/2^{m}$.

\vskip .10in
The complementary event consists of the following $m+1$ mutually exclusive events:  First, there is an integer $k$ with $1 \leq i \leq m$
such that for $i \leq k$ each $d_{i}$ is even and for all $j > k$ each $d_{j}$ is odd; and, second, all values of $d_{k}$ are odd.  Each of these
events has probability $1/2^{m}$, hence the union has probability $(m+1)/2^{m}$, so then $p(m)$ has probability $1-(m+1)/2^{m}$, as claimed.
\end{proof}

\vskip .10in
\begin{remark}\rm
In the statement of Lemma \ref{lem: repetitions in S}, we assumed that we were
given the two elements in $S$ which form a repetition, meaning they are equivalent
$\text{\rm \, mod \,} N$.  In other words, the analysis in Lemma
\ref{lem: repetitions in S} begins with \eqref{repetition}.
Since $S$ has $O(\log _{2}N)$ elements, the straightforward, exhaustive search in $S$ to
determine if \eqref{repetition} occurs takes at most $O((\log_2 N)^{2})$ deterministic steps.
\end{remark}




\vskip .10in
\begin{lemma}\label{claim: order of a cyclic subgroup of Z}
Let $G$ be a finite cyclic group of even order $n$, which we write as
$n=2^{c}m$ where $c>0$ and $m$ is an odd integer. Let $u$ be a generator of $G$.  Then for $1\leq d\leq 2^{c}m$ we have the following statements.
\begin{enumerate}
\item If $d$ is odd, then $\mathrm{ord}_G(u^{d})=2^{c}k^{\prime }$, where $k^{\prime }$ is odd.
\item If $d$ is even, which we write as $d=2^{v}d^{\prime }$ with $d^{\prime }$ odd, then $\mathrm{ord}_G(u^{d})=2^{c-t}m^{\prime }$, where $t=\min (c,v)>0$ and $m^{\prime }$ is odd.
\end{enumerate}
\end{lemma}

\vskip .10in
\begin{proof}
The assertion follows immediately from the elementary observation that $\mathrm{ord}_G(u^{d})=n/\gcd(n,d). $
\end{proof}

\vskip .10in
\begin{lemma}\label{lem: reducing to odd order}
For any $a \in \mathbb{Z}_N^{\ast}$, the largest power of $2$ that divides $\mathrm{ord}_N(a)$ is less than $\log _{2}N$. Therefore, if $M=[\log _{2}N]+1$, then the order of $b = a^{2^{M}}$ must be odd.
\end{lemma}
\begin{proof}
As above, let $g_N(a)=(u_{1}^{d_{1}},\ldots, u_{m}^{d_{m}})$, from which
we have that
$$
\mathrm{ord}_N(a)=\text{lcm}(\mathrm{ord}_{p_1}(u_{1}^{d_{1}}), \ldots, \mathrm{ord}_{p_m}(u_{m}^{d_{m}})) = \text{lcm}(2^{c_1}p_1', \ldots, 2^{c_m}p_m').
$$

\vskip .10in
Then by Lemma \ref{claim: order of a cyclic subgroup of Z} the largest power of $2$ that divides $\mathrm{ord}_N(a)$ is less than or equal to $\max \{c_{1},\ldots, c_m\}$ which is
necessarily less than $\log _{2}N$ because
$$
\log _{2}N > \log _{2}\left( \prod_{i=1}^{m} 2^{c_i}p_i'\right)= \sum_{i=1}^{m}\left( c_i +  \log _{2} p_i'\right) > \max \{c_{1},\ldots, c_m\}.
$$

\vskip .10in
Let $r=\mathrm{ord}_N(a)$, and write $r=2^{v}r^{\prime }$ where $r'$ is odd.
 Since $r<N$, it follows that $v<M$.  Let $\mathrm{ord}_N(a^{2^{M}})=2^{e}t$ where $t$ is odd and $e\geq 0$.  Then $r \mid  2^{M+e}t$; consequently, $r^{\prime }\mid t$, so we
 can factor $t$ as $t=r't'$ where $t'$ is odd.   Then
$$
a^{2^{M}t} = a^{2^{v}2^{M-v}r'  t'} = (a^{2^{v}r'})^{2^{M-v}t'}
\equiv 1 \textrm{\rm \,mod\,} N.
$$
Thus, $\mathrm{ord}_N(a^{2^{M}})$ divides $t$, which is odd,
so then $\mathrm{ord}_N(a^{2^{M}})$ itself is odd, as claimed.
\end{proof}

\vskip .10in
\begin{proposition}\label{lem: factoring to order finding}
Set $M=[\log _{2}N]+1$.  For any $a \in \mathbb{Z}_N^{\ast}$, let $r_b = \mathrm{ord}_N(a^{2^{M}})$. If $r_b$ is known, we can compute $r_{a}=\mathrm{ord}_N(a)$
in at most $O(\log_2 N)$ deterministic steps.  Furthermore, $r_a$ is even with probability at least $p(m) = 1 - (m+1)/2^m$ in which case $x = a^{r_a/2}$ satisfies
$$
x^{2} \equiv 1 \textrm{\rm \,mod\,} N
\,\,\,\,\,\,\,\,
\textrm{\it and}
\,\,\,\,\,\,\,\,
 x\not\equiv \pm 1 \textrm{\rm \,mod\,} N.
$$
\end{proposition}
\begin{proof}
By Lemma \ref{lem: reducing to odd order}, $r_b$  is odd.  Then $\mathrm{ord}_N(a)$
is equal to $2^{k}r_b$, where $k$ is the smallest exponent such that
$a^{2^{k}r_b}\equiv 1 \textrm{\rm \,mod\,} N$.  Necessarily, we have that $k \leq M$. Given $r_b$, we can compute such $k$ in at most $O(\log_2 N)$ steps. Let $ g_N(a)=(u_{1}^{d_{1}},\dots, u_{m}^{d_{m}}).$ Proposition \ref{prop: main elementary NT} shows that $k > 0$, i.e. $\mathrm{ord}_N(a)$ is even if there exist $i,j$ with $1\leq i<j \leq m$ such that $d_{i}$ is odd and $d_{j}$ is even . According to the proof of Lemma \ref{lem: repetitions in S}, the probability of this is at least $p(m)$. The statement now follows from Proposition \ref{prop: main elementary NT}.
\end{proof}


\vskip .10in
\section{Proof of Theorem \ref{thm: main}} \label{sec: proof of main thm}
Let $N\geq 2$ be a fixed positive integer, and let $b$ be an integer which is relatively prime to $N$ and of \emph{odd} order $r$ in $\mathbb{Z}_N$.
The proof that we can determine $r$ in at most $O((\log_2 N)^2)$
diffusion steps runs as follows.

\vskip .10in
\begin{enumerate}
\item
We construct an appropriate weighted Cayley graph $X_{r, S}$ with $r$ vertices
and study the half-lazy random walk $\{p ^{X_{r,S}}_k\}_{k=0}^\infty$
on this graph in $n$ steps.  The construction of the graph is undertaken without
the explicit knowledge of $r$.
\item
By using the bounds for the Korobov-type exponential sums as proved in \cite{KM12},
we will deduce an upper bound for the second largest eigenvalue
$\lambda^{X_{r,S}}_\ast$ of the half-lazy random walk on $X_{r,S}$.
We note here that results related to those from \cite{KM12} are given in
\cite{LLW98}, \cite{Mo09} and \cite{Va09}.
\item
We determine the number of diffusion steps $n$ so that
$\lambda^{X_{r,S}}_\ast < N^{-2}$.  In doing so, the integer $r$ can be characterized as
the integer closest to $1/p_n^{X_{r,S}}(e)$, where $e=0$ is the starting vertex of the graph $X_{r,S}$ in additive notation.
\end{enumerate}

\vskip .10in
As it will be shown, $n$ can be taken to be the smallest integer
bigger than $4\log N (\lfloor\log_2 N\rfloor+2)$.

\vskip .10in
\subsection{Graph construction in multiplicative notation}
For a fixed $b\in \mathbb{Z}^\ast_N$ of odd order $r$, let $G_{N,b} = \langle b \rangle \subseteq \mathbb{Z}^\ast_N$ be the subgroup of $\mathbb{Z}^\ast_N$ generated by $b$.
The elements of $G_{N,b}$ are the classes of $b^k \textrm{\rm \,mod\,} N$
for $k=0,1, \dots, r-1$.
Take $S_{N,b}=\{b^{\pm 2^{t}}:$ $t=0,\ldots ,M\}$ with $M =\lfloor \log_2 N\rfloor+1$. Define the weight function $\alpha_{N,b}$ by
$$
\alpha_{N,b}(b^{ 2^{t}}):= |\{l\in\{0,\ldots,M\} \colon b^{2^{t}} \equiv b^{2^{l}}
\textrm{\rm \,mod\,} N \textrm{\rm \, or \,}
 b^{ 2^{t}} \equiv b^{-2^{l}}\textrm{\rm \,mod\,} N\}|,
$$
where $|A|$ denotes the number of elements of a finite set $A$. It is immediate that $\alpha_{N,b}(b^{ 2^{t}})=\alpha_{N,b}(b^{- 2^{t}})$, for $t=0,\ldots ,M$.  Therefore,
$\alpha_{N,b}$ can be used as the weight function in the construction of the weighted Cayley graph in the multiplicative notation, as in \cite{Ba79}.

\vskip .10in
Let $X_{N,b} = \text{Cay}(G_{N,b}, S_{N,b},  \alpha_{N,b})$ be the weighted Cayley graph of $G_{N,b}$ with respect to $S_{N,b}$ and the weight function  $\alpha_{N,b}$. It is immediate that $X_{N,b}$ has $r$ vertices and it is a regular graph of degree $2(M+1)$.

\vskip .10in
{\it It is important to note that in the construction of $X_{N,b}$ we do not know
the value of $r$.  Indeed, all that is required is the value of $N$ since we begin with
one point $b$ and, recursively, let the diffusion process develop in $2(M+1)$ possible directions from any given point.}

\vskip .10in
From the beginning, we do not know the entire graph.
Nevertheless, since diffusion is local in nature, this allows us to build $X_{N,b}$
one diffusion step at a time. Our main theorem states that after a number of steps which is polynomial in $\log_2 N$ we will have enough information to approximate the number of vertices of $X_{N,b}$, and thus the order of $b$. What makes the process effective is
the fact that diffusion occurs simultaneously at all constructed vertices, which provides
 some form of parallel computation.

\subsection{An equivalent description of the graph}

In this section we will describe the graph $X_{N,b}$
in additive notation, which we find to be more amenable for describing the bounds for the eigenvalues.
Since $b$ is fixed, we will simplify the notation by omitting the index $b$ and emphasizing the dependence upon $r$.

\vskip .10in
\it
It is important to emphasize that, ultimately, we will use diffusion to compute $r$.  The only information we will
use about $r$ itself is that $0 < r \leq N$ and that $r$ is the order of an element $b$ modulo $N$. Nonetheless, this
section is provided as a notational aid in our proofs of Theorem \ref{thm: main} and Theorem \ref{thm: integer fact}.
\rm
\vskip .10in

Let $C_r=\{0,\ldots ,r-1\}$ denote the additive group of integers modulo $r$, and set
$$
S=\{\pm 2^{j}:j=0,\ldots ,M\}
\,\,\,\,\,\,\,
\textrm{\rm with}
\,\,\,\,\,\,\,
M=\lfloor \log_2 N\rfloor+1.
$$
Note that $S\subseteq C_{r}$ is a symmetric set which generates $C_r$, 
and that the numbers $\pm 2^j$ for $j\in\{0,\ldots ,M\}$ are not necessarily distinct modulo $r$.  We define the weight function $\alpha\colon S \to \mathbb{R}^{>0}$ as
follows. For $2^j \in S$, we let
$$
\alpha(2^j):= |\{l\in\{0,\ldots,M\} \colon 2^j \equiv 2^l\textrm{\rm \,mod\,} r
 \textrm{\rm \, or \,}  2^j \equiv -2^l\textrm{\rm \,mod\,} r\}|.
$$
The congruence $\pm2^j \equiv -2^l\textrm{\rm \,mod\,} r$ is equivalent
 to $\pm2^j \equiv 2^l\textrm{\rm \,mod\,} r$.  Hence, $\alpha(-2^j)=\alpha(2^j)$,
so it can be used as the weight function in the construction of the weighted
Cayley graph with respect to $S$.

\vskip .10in
{\it Since $b$ has order $r$ modulo $N$, the weight function $\alpha$ is equal to
the weight function $\alpha_{N,b}$.  As a result, the values of $\alpha$ can be
determined by computing powers of $b$ modulo $N$ where the value of $r$ is unknown.}

\vskip .10in
Let $X_{r,S} = \text{Cay}(C_r, S, \alpha)$ be the weighted Cayley graph of $C_r$ with respect to $S$ and $\alpha$. The graph $X_{r,S}$ has $r$ vertices, and it is regular of degree $\vert S \vert = 2(M+1)$. In case all numbers $\pm 2^j$ with $j\in\{0,\ldots ,M\}$ are distinct modulo $r$, the graph  $X_{r,S}$ is the Cayley graph of $C_r$ with respect to $S$. If there are repetitions modulo $r$ in the sequence $\pm 2^j$ with $j\in\{0,\ldots ,M\}$, then we can view $X_{r,S}$ in such a way that the vertex $x\in C_r$ is connected to the vertex $y\in C_r$ with possibly more than one edge; the number of edges connecting $x$ and $y$ being the number of elements of the set $\{l\in\{0,\ldots,M\} \colon x-y \equiv 2^l\mod r \text{  or  }  x-y \equiv -2^l\mod r\}$.

\vskip .10in
For this choice of $S$, the eigenvalues $\eta _{k}$ for $k=0,\ldots,r-1$ of the adjacency matrix for the graph are known.  Specifically, $\eta_0 = 2(M+1)$ and
\begin{align*}
    \eta_{k} = \sum_{x\in S}\alpha(x)e^{\frac{2\pi i}{r}kx} = \sum_{j=0}^{M}e^{\frac{2\pi i}{r}k2 ^{j}} +\sum_{j=0}^{M}e^{-\frac{2\pi i}{r}k2 ^{j}} \quad \text{for each}\quad 1\leq k\leq r-1.
\end{align*}

\vskip .10in
Thus, the eigenvalues of the half-lazy walk matrix $W_{r,S}$ of $X_{r,S}$ are
$$
\lambda^{X_{r,S}}_{k}=\frac{1}{2}\left(1+\frac{\eta _{k}}{2(M+1) }\right)
\,\,\,\,\,\,\,
\textrm{\rm for}
\,\,\,\,\,\,\,
k=0, 1,\dots,r-1.
$$
Proposition \ref{prop: bound for the heat flow} implies that for any vertex $x$
we have 
\begin{equation}\label{eq:bound for probabilities}
\left\vert p^{X_{r,S}}_n(x) - \frac{1}{r} \right\vert \leq (\lambda^{X_{r,S}}_\ast)^n,
\end{equation}
where, as above, $\lambda^{X_{r,S}}_\ast$ is the largest eigenvalue of $W_{r,S}$
less than 1.


\subsection{Eigenvalue bounds}

Our next task is to find an upper bound independent of $r$ for the quantity
\begin{equation}\label{eq: general eta k multiset}
\frac{1}{2(M+1)}\max_{k\in\{1,\ldots r-1\}}\left|\eta_k\right|=\frac{1}{2(M+1)}\max_{k\in\{1,\ldots r-1\}}\left|\sum_{j=0}^{M}e^{\frac{2\pi i}{r}k2 ^{j}} + \sum_{j=0}^{M}e^{-\frac{2\pi i}{r}k2 ^{j}} \right|.
\end{equation}
Let $k\in\{1,\ldots,r-1\}$ be arbitrary, and let $d=\mathrm{gcd}(k,r)$.
Set $k_1=k/d$ and $r_1=r/d$.  Then, we have that
\begin{equation}\label{eq: sum up to M}
\sum_{j=0}^{M}e^{\frac{2\pi i}{r}k \cdot 2^j}=\sum_{j=0}^{M}e^{\frac{2\pi i}{r_1}k_1 \cdot 2^j}.
\end{equation}
Since $r_1$ and $k_1$ are relatively prime, $e^{\frac{2\pi i}{r_1}k_1}$ is a primitive
$r_1$-th root of unity. Let $\tau_1=\mathrm{ord}_{r_1} 2$.
In order to bound \eqref{eq: sum up to M}, we consider the following three cases.

\vskip .10in
\begin{itemize}
\item[(i)] {\bf Assume $r_{1} \geq 4$ and $M\geq \tau_1-1$.} Since $M+1\geq \tau_1$,
 we can write $M+1=q\tau_1 +s$ for $q\geq 1$ and some $0\leq s\leq\tau_1-1$.
 Equivalently, $M=q\tau_1 + s_1$, for $q\geq 1$ and $-1\leq s_1\leq
 \tau_1-2$.  Then, we have that
$$
\sum_{j=0}^{M}e^{\frac{2\pi i}{r_1}k_1 \cdot 2^j} =q \sum_{j=0}^{\tau_1-1}e^{\frac{2\pi i}{r_1}k_1 \cdot 2^j}+\sum_{j=0}^{s_1}e^{\frac{2\pi i}{r_1}k_1 \cdot 2^j},
$$
where, in the case that $s_1=-1$ the sum on the right-hand side is taken to be zero.
The second statement of Corollary 1 of \cite{KM12}, in our notation, states that when $r_1 >3$ we have the bound
$$
\max_{(k_1,r_1)=1}\left|\sum_{j=1}^{\tau_1}e^{\frac{2\pi i}{r_1}k_1 \cdot 2^j} \right|<\tau_1-1,
$$
where the maximum is taken over all pairs of coprime $k_1$ and $r_1$. Since  $\tau_1=\mathrm{ord}_{r_1} 2$, it is immediate that
$$
\sum\limits_{j=1}^{\tau_1}e^{\frac{2\pi i}{r_1}k_1 \cdot 2^j} = \sum\limits_{j=0}^{\tau_1-1}e^{\frac{2\pi i}{r_1}k_1 \cdot 2^j}.
$$
Therefore,
\begin{align*}
\left|\sum_{j=0}^{M}e^{\frac{2\pi i}{r_1}k_1 \cdot 2^j}\right|&\leq q \left| \sum_{j=0}^{\tau_1-1}e^{\frac{2\pi i}{r_1}k_1 \cdot 2^j}\right|+ \left| \sum_{j=0}^{s_1}e^{\frac{2\pi i}{r_1}k_1 \cdot 2^j}\right|\\& \leq q(\tau_1-1) +s_1+1\\&=M+1-q\leq M.
\end{align*}

\vskip .10in
\item[(ii)] {\bf Assume $r_{1} \geq 4$ and $M < \tau_1-1$.}
    Since $r_{1}$ is odd, we actually have that $r_{1} \geq 5$.  In this case we proceed analogously as in the proof of \cite{KM12}, Theorem 2. This approach
    is justified, because in the notation of \cite{KM12}, $\mathcal{L}=\lfloor\log_2 r_1\rfloor \leq \lfloor\log_2 r\rfloor \leq \lfloor\log_2 N\rfloor = M-1 $, so there are sufficiently many summands in the series so that the results of \cite{KM12}, Lemma 3, apply.

    More precisely, Lemma 3 from \cite{KM12}, claims that if $(a,q)=1$, $q>3$, then for every integer $m\geq 0$ there exist an integer $\ell$ with $m<\ell\leq \lfloor \log_2 q\rfloor +1+m$ and such that
$$
\left|\exp\left(\frac{2\pi i}{q} a 2^{\ell}\right) -1 \right|<\left|\exp\left(\frac{2\pi i}{q} a 2^{\ell -1}\right) -1 \right|.
$$
Taking $q=r_1\geq 5$, $a=k_1$ and $m=0$, we conclude that there exist an integer $\ell_0\in\{1,\ldots,M\}$ such that
\begin{equation*}\label{eq: minimal ell}
  \left|\exp\left(\frac{2\pi i}{r_1}k_1 2^{\ell_0}\right) -1 \right| < \left|\exp\left(\frac{2\pi i}{r_1}k_1 2^{\ell_0-1}\right) -1 \right|.
\end{equation*}


\vskip .10in

Therefore,
\begin{align*}
\left|\exp\left(\frac{2\pi i}{r_1}k_1 2^{\ell_0-1}\right)+\exp\left(\frac{2\pi i}{r_1}k_1 2^{\ell_0}\right) \right|&\leq \left|\exp\left(\frac{2\pi i}{r_1}k_1 2^{\ell_0-1}\right)+1 \right|\\&= \frac{\left|\exp\left(\frac{2\pi i}{r_1}k_1 2^{\ell_0}\right) -1 \right|}{\left|\exp\left(\frac{2\pi i}{r_1}k_1 2^{\ell_0-1}\right) -1 \right|} <1.
\end{align*}

\vskip .10in
This shows that
\begin{align}\notag
\left|\sum_{j=0}^{M}e^{\frac{2\pi i}{r_1}k_1 \cdot 2^j}\right| &= \left|\exp\left(\frac{2\pi i}{r_1}k_1 2^{\ell_0-1}\right)+\exp\left(\frac{2\pi i}{r_1}k_1 2^{\ell_0}\right) \right|\\& + \sum_{j\in\{0,\ldots,M\} \setminus\{\ell_0-1,\ell_0\}}\left|e^{\frac{2\pi i}{r_1}k_1 \cdot 2^j}\right|< 1+M-1=M.\label{eq: bound by M}
\end{align}
\item[(iii)] {\bf Assume $r_{1} =3$.}
In this case $r=3d$.  Since $k_{1}$ and $r_{1}$ are relatively prime, $k_{1} \in\{1,2\}$, so then $k=k_1d<3d$. Therefore, we need to find the upper bound for the absolute value
of the sums
$$
\sum_{j=0}^{M}e^{\frac{2\pi i}{3} \cdot 2^j}\quad \text{or} \quad \sum_{j=0}^{M}e^{\frac{4\pi i}{3}\cdot 2^j}.
$$
Trivially, both sums are bounded by $M$.  Indeed, all terms in either sum are
cube roots of unity.  In both case, the terms corresponding to $j=0$ and $j=1$ add
to give $1/2 \pm i\sqrt{3}/2$, respectively, which has absolute value equal to one.
The remaining $M-1$ terms have absolutely value equal to one, from which the bound of
$M$ for each series is obtained.
\end{itemize}

\vskip .10in
Since $e^{-\frac{2\pi i}{r}k\cdot 2 ^{j}}=e^{\frac{2\pi i}{r}(r-k)\cdot 2 ^{j}}$, the inequality \eqref{eq: bound by M} holds true for the negative exponents as well. Hence the bound for \eqref{eq: general eta k multiset} becomes
$$
\frac{1}{2(M+1)}\max_{k\in\{1,\ldots r-1\}}|\eta_k|<\frac{2M}{2(M+1)}= 1-\frac{1}{M+1}.
$$

\vskip .10in
In summary, we have proved that the largest eigenvalue of the half-lazy
random walk matrix $W_{r,S}$ on $G_{r,S}$ satisfies the bound
\begin{equation}\label{eq: bound for largest eigenv}
\lambda^{X_{r,S}}_\ast \leq \frac{1}{2} \left( 1+1-\frac{1}{M+1} \right)=1-\frac{1}{2(M+1)}.
\end{equation}

\subsection{Counting the number of diffusion steps}\label{counting_heat_steps}
It is now left to establish the number of diffusion steps needed to determine $r$ by performing the half-lazy random walk on the weighted graph $X_{r, S}$, constructed above, by starting at vertex $e = 0$.

\vskip .10in
By combining \eqref{eq: bound for largest eigenv} with \eqref{eq:bound for probabilities} we deduce, for any positive integer $n$, that
$$ \left\vert p^{X_{r,S}}_n(e) - \frac{1}{r} \right\vert \leq \left(1-\frac{1}{2(M+1)}\right)^n.
$$
We know that $p_n^{X_{r,S}}(e)$ converges to $1/r$ as $n$ tends to infinity.
For any two distinct positive integers $m_{1}$ and $m_{2}$ less than $N$,
the smallest distance between their reciprocals $1/m_{1}$ and $1/m_{2}$ is bounded from
below by
$$
\frac{1}{N} - \frac{1}{N-1} = \frac{1}{N(N-1)} \geq \frac{1}{N^2}.
$$
Thus, if we have $p_n^{X_{r,S}}(e) = 1/r$ within an error of $1/N^2$, we have determined $r$.
Therefore, the smallest number of diffusion steps needed to determine $r$ is bounded
from above by the smallest positive integer $n$ for which
$$
\left(1-\frac{1}{2(M+1)}\right)^n<\frac{1}{N^2}.
$$
Since $N \geq 2$, we have that $0<1/(2M)<1$ so then
$$
-\log\left(1-\frac{1}{2(M+1)}\right)=\sum_{\ell=1}^{\infty}
\frac{1}{\ell(2(M+1))^{\ell}}>\frac{1}{2(M+1)}.
$$
Choose $n$ to be the smallest integer for which
$$
n >  4 (M+1) \log N = 2 \log N (\lfloor\log_2 N \rfloor+2).
$$
Then, the inequality
$$
\left\vert p^{X_{r,S}}_n(e) - \frac{1}{r} \right\vert < \frac{1}{N^2}
$$
holds true, and then we can compute the integer $r$ uniquely after $n$ diffusion steps, followed by one measurement (of the ``heat" at the starting point $e$).

\vskip .10in
\section{Proof of Theorem \ref{thm: integer fact}} \label{sec: Proof of Thm integer fact}
\noindent
Our algorithm receives as input a positive integer $N$
which is assumed to be neither prime nor the power of a prime. The algorithm returns a divisor $d$ of $N$,
with probability at least $p(m)$,  and it runs as follows:
\begin{enumerate}
\item Select $a \in \mathbb{Z}_N=\{1,\ldots,N\}$ uniformly at random.  \\
Compute $d = \gcd(a, N)$. \\
\phantom{aaa} If $1< d < N$, return $d$. \\
\phantom{aaa} Else go to step 2.

\item Compute the set $S=\{a^{ \pm 2^{t}} \textrm{\rm \,mod\,} N:
t=0,\ldots ,M\}$ where  $M=\lfloor \log_2 N\rfloor+1$.

\item If there are repetitions in $S$, say $a^{2^{l}}=a^{\pm 2^{l^{\prime }}}$ with $l>l^{\prime }$, do: \\
\phantom{aaa} Set $q=2^{l-l^{\prime }} \pm 1$. \\
\phantom{aaa} Compute the smallest integer $s \geq 0$ such that
$ a^{2^{s}q}\equiv 1 \textrm{\rm \,mod\,} N$. \\
\phantom{aaa} If $s> 0$, compute $d = \gcd(a^{2^{s-1}q} - 1, N)$ \\
\phantom{aaa} \phantom{aaa} If $1< d < N$, return $d$. \\
\phantom{aaa} \phantom{aaa} Else terminate (with no answer).  \\
\phantom{aaa} Else terminate (with no answer).  \\
Else go to step 4.

\item Set $b = a^{2^M}$. \\
Run the diffusion computer algorithm to determine the order of $b$ modulo $N$.\\
Set $r_{b} = \text{ord}_N(b)$.

\item Compute the smallest integer $k \geq 0$ such that  $ a^{2^{k}r_b}\equiv 1 \textrm{\rm \,mod\,} N$. \\
Set $r_a = 2^k r_b$ which is the order of $a$ modulo $N$. \\
If $r_a$ is even, compute $d = \gcd(a^{r_a/2} - 1, N)$ \\
\phantom{aaa} If  $1< d < N$, return $d$. \\
\phantom{aaa} Else terminate (with no answer). \\
Else terminate (with no answer).
\end{enumerate}

\vskip .10in
The Euclidean algorithm determines the greatest common divisor of two numbers in $\mathbb{Z}_N$ using at most $O((\log N)^2)$ deterministic steps.  With this, we
can bound the complexity of each of the above steps as follows.

\begin{itemize}
    \item[ ] Step 1: This steps uses the Euclidean algorithm to find $\gcd(a,N)$, so it runs in deterministic time $O((\log N)^2)$.
    \item[] Step 2: Using the method of repeated squaring, we can compute the set $S$ in at most $O(2M) = O(\lfloor \log_2 N \rfloor + 1)$ deterministic steps.
    \item[] Step 3: Checking for repetitions in $S$ requires $O(\log_2 N)$ steps. By Lemma \ref{lem: repetitions in S}, we can compute the smallest integer $s \geq 0$ such that $a^{2^{s}q}\equiv 1 \textrm{\rm \,mod\,} N$ in $O(\log_2 N)$ deterministic steps. If $s>0$, computing $\gcd(a^{2^{s-1}q} - 1, N)$ takes $O((\log N)^2)$ more
        deterministic steps.
    \item[] Step 4: Theorem \ref{thm: main} states that using a diffusion computer allows us to compute $r_b$ in at most $O((\log_2 N)^2)$ diffusion
         steps.
    \item[] Step 5: By Lemma \ref{lem: factoring to order finding}, we can compute $r_a$, the order of $a$, in $O(\log_2 N)$ deterministic steps. If $r_a$ is even, computing $\gcd(a^{r_a/2} - 1, N)$ takes $O((\log N)^2)$ more deterministic steps.
\end{itemize}

The algorithm runs Steps $1$ and $2$, and then runs either Step $3$ or Steps $4$ and $5$. Therefore, the algorithm runs in at most $O((\log N)^2)+ O(\lfloor \log_2 N \rfloor + 1) + O(\log_2 N) +O((\log N)^2)$ determistic steps plus at most $O((\log_2 N)^2)$ diffusion steps.

\vskip .10in
The algorithm is successful if it returns a nontrivial factor $d$ of $N$. Both Steps $3$ and $5$ have a success probability of at least $p(m) = 1 - (m+1)/2^m$, by Lemmas
\ref{lem: repetitions in S} and \ref{lem: factoring to order finding}. Hence, whether the algorithm runs Step $3$ or Steps $4$ and $5$, the success probability is at least $p(m)$.

\section{Examples}  \label{Ejemplos}

The following examples illustrate the above described algorithm.

\subsection{Example 1: \it N = 33}

\begin{enumerate}
\item[Step 1.] We choose $a=5$ which is relatively prime to $33$.
\item[Step 2.] $M=\lfloor{\log_2(N)}\rfloor+1=6$, and $S=\{5^{2^0}=5,5^{2^1}\equiv 25,\ldots,5^{2^5}\equiv 25,5^{-2^1}\equiv 20\ldots\} \textrm{\rm \,mod\,} 33$.
\item[Step 3.] A repetition in $S$ is detected: $5^{2^5-2^1}\equiv 1 \textrm{\rm \,mod\,} 33$, so then $s=1$ and
$d=\gcd(5^{2^{0}\cdot 15}-1,33)=11$.
\end{enumerate}

Therefore, $11$ divides $N$, from which we obtain that $N=3\times11$.

\subsection{Example 2: \it N = 1363}

\begin{enumerate}
    \item[Step 1.] We choose $a=991$ which is relatively prime to $1363$.
    \item[Step 2.] $M=\lfloor{\log_2(N)}\rfloor+1=11$, and $S=\{a^{\pm 2^{t}}: t=0,\ldots,11\}=\{991^{2^0}=991,991^{2^1}=721,\ldots,991^{2^{11}}=944,991^{-2^{1}}=905,\ldots\}\textrm{\rm \,mod\,} 1363$. There are no repetitions in $S$. Therefore we go to Step 4.
    \item[Step 4.]We set $b=991^{2^{11}}\equiv 944\textrm{\rm \,mod\,} 1363$ and calculate the set $S=\{b^{\pm 2^{t}}: t=0,\ldots,11\}$. 
    We run the diffusion computer in order to determine the order $r_b$ of $b=944$.

  {\bf Note:} In order to get an estimate for $1/r_b$ with an error less than $1/N^2\approx5.38\times 10^{-7}$, the diffusion computer requires at least $\lfloor 4(M+1)\log(N) \rfloor +1=347$ diffusion steps.
   As we will describe below, after $36$ diffusion steps, consisting of $n=25$ iterations
   of the diffusion process and $11$ measurements, we were able to conclude that $r_b = 161$.

    \item[Step 5.] The smallest non negative integer $k$ such that $991^{2^k \times 161} \equiv 1 \textrm{\rm \,mod\,} 1363$ turns out to be $1$. We conclude that $r_a=322$. So then we computed $\gcd(991^{161}-1,1363) = 47$. Thus,
        $47$ divides $N$, from which we obtain that $N=47\times 29$.
\end{enumerate}

\vskip .10in
Note that since $47$ and $29$ are prime and odd, then $r_{a}$ must divide
$(47-1)(29-1) = 1288$.  Indeed, we have that, as expected, $1288 = 322 \cdot 4$. 

\vskip .10in
Let us now discuss the details behind Step 4.  After iterating the diffusion process $n=25$
times, we then measured the probability distribution $p_{25}(v)$ for the $11$ values
of $v$ which are the vertices on the graph corresponding to $S$.  
As it turns out, for
each such $v$ we had that $1/160 < p_{25}(v) < 1/162$. This narrows the possibilities for the order of $b$ to just three integer values.  
By trying every one of these values we then confirmed that, indeed, $r_{b}=161$.  

\vskip .10in
Though not needed, the \it Maple \rm code in the Appendix produces the values of $p_{25}(v)$ for all $v$. A histogram for the reciprocals of these probabilities is presented in Figure 1.
\begin{figure}[h!]
	\centering
	\includegraphics[scale=0.6]{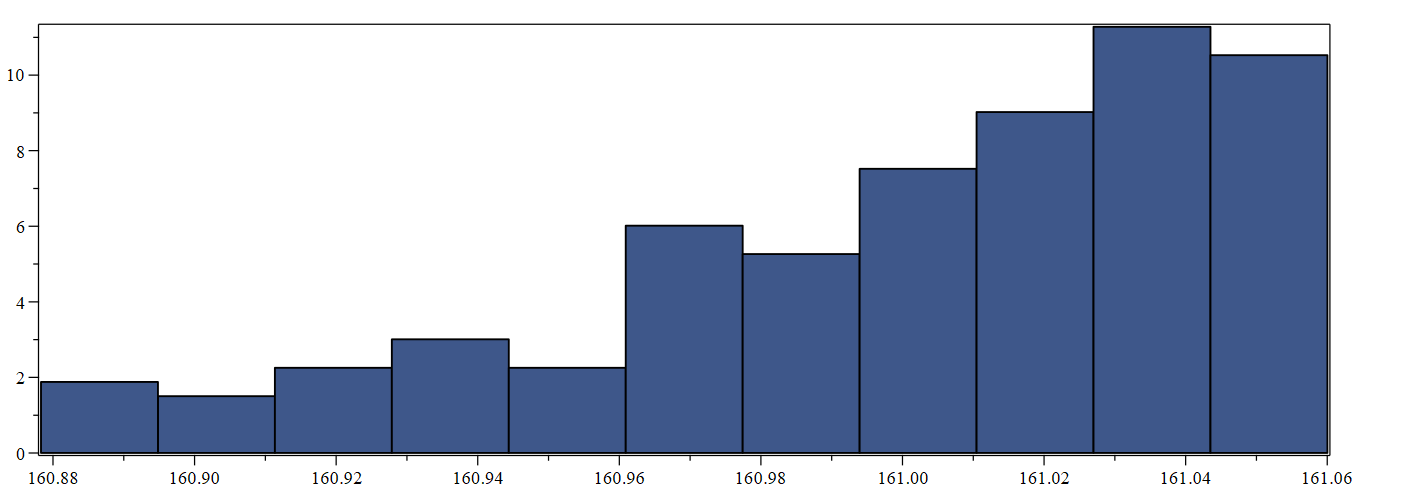}
 \caption{Histogram for the reciprocals of probabilities after 36 diffusion steps}
\end{figure}

\vskip .10in
In the next section we discuss a way in which the methodology of Example 2 can be
encoded, possibly yielding a faster algorithm.

\section{Concluding remarks}

\subsection{Reducing the number of diffusion steps}

Note that the bound in Proposition \ref{prop: bound for the heat flow} is somewhat crude since it used inequality \eqref{eq: l2 bound for distance}.
More specifically, it is possible that there are additional cancellations in the exponential sums appearing in the spectral expansion of the heat kernel for the half-lazy
random walk.  If so, then one would need fewer than the number of diffusion steps given by \eqref{diffusion_bound}  in Theorem \ref{thm: main}.  The discussion provides a guide by which
the implementation of the diffusion computation may yield results before reaching  the number of steps stated in \eqref{diffusion_bound}.

Let us suppose that after $m$ diffusion steps we have that
\begin{equation}\label{general_bound}
\left\vert p_m(v) - \frac{1}{r} \right\vert \leq A_{m}
\end{equation}
for any vertex $v$ and some constant $A_{m}$.  This is equivalent to
$$
p_m(v)  - A_{m} \leq \frac{1}{r} \leq p_m(v)  + A_{m}.
$$
Trivially, one has that $1/N \leq 1/r \leq 1$.  
Let $\mathcal{S}$ denote any subset of the vertices of $X$.  We
now can conclude that
\begin{equation}\label{interval_bounds}
\max\left(1/N, \max_{v\in \mathcal{S}} \left(p_m(v)  - A_{m}\right)\right)
\leq \frac{1}{r} \leq \min\left(1,\min\limits_{v\in \mathcal{S}} \left(p_m(v)  + A_{m}\right)\right).
\end{equation}
Proposition \ref{prop: bound for the heat flow} shows that the worst-case
scenario happens when $A_{m} = \lambda_{1}^{m}$.  However, it may be the case that additional
cancellations occur in the spectral expansion  (\ref{eq: p_n computation}) and in the exponential sums defining $\lambda_1$.

\vskip .10in
In general, let us suppose that, for whatever reason, after $m$ diffusion steps we have the bound \eqref{general_bound} for some $A_{m}$.  Then the order $r$ is amongst the integers $h$ such
that $1/h$ lies in the interval
\begin{equation}\label{interval}
\mathcal{I}_{\mathcal{S}} := \left(\max\left(1/N, \max_{v\in \mathcal{S}} \left(p_m(v)  - A_{m}\right)\right), \min\left(1,\min\limits_{v\in \mathcal{S}} \left(p_m(v)  + A_{m}\right) \right)\right)
\bigcap \left\{1/h \in \mathbf{Q} : h \in \mathbf{Z}\right\}.
\end{equation}
Theorem \ref{thm: main} amounts to saying that by taking $m=\lfloor 4(M+1)\log(N) \rfloor +1=O((\log N)^{2})$,
there is only one integer, namely the order $r$, such that $1/r$ lies in the interval
\eqref{interval}.

\vskip .10in
Certainly, for smaller $m$, and in the presence of an improved bound
for $A_{m}$, the interval \eqref{interval} could contain a small 
number of entries.
One could stop the implementation
of further diffusion steps and then check one by one each entry of $\mathcal{I}_{S}$ to determine which
value yields the sought-for order.  In other words, one can reduce
the number of iterations of the symmetric matrix $W$ from the bound stated in \eqref{diffusion_bound} to $m$, but then increase
the number
of measurements from one to $\vert\mathcal{I}_\mathcal{S}\vert$, where $\vert\mathcal{I}_\mathcal{S}\vert$ is the cardinality
of $\mathcal{I}_\mathcal{S}$.  If it takes $t$ steps on a digital computer to determine if a given
integer $h$ is the sought-for order, one would increase the number of digital steps
by $t\cdot \vert\mathcal{I}_{\mathcal{S}}\vert$.  For example, if for some constant $c > 0$ we take
$$
m = 2\log( (\log N)/c)(\lfloor \log_{2} N \rfloor+ 2),
$$
then, as in section \ref{counting_heat_steps}, we get that
$$
A_{m} \leq \left(1 - \frac{1}{2(M+1)}\right)^{m} \leq \left(\frac{c}{(\log N)}\right)^{2}.
$$
Let us rewrite the interval in \eqref{interval_bounds} as
$$
L_{m} \leq 1/r \leq U_{m}.
$$
then
$$
\vert\mathcal{I}_{\mathcal{S}}\vert \leq \frac{1}{L_{m}} - \frac{1}{U_{m}} =
\frac{U_{m}-L_{m}}{L_{m}U_{m}}
$$
and $\vert\mathcal{I}_{\mathcal{S}}\vert$ could be as small as $O((\log N)^{k})$
for some constant $k \geq 2$.  

\vskip .10in
In other words, there may be circumstances under which the number of diffusion steps could
be reduced to $O(\log N)$, with an effective constant, while the number of digital steps
would become $O((\log N)^{k})$, thus not significantly changing the complexity of the number
of digital steps.  
This discussion leads to an interesting optimization problem, we which will
leave for a future study.

\subsection{Searching for additional repetitions}

The algorithm which proves Theorem \ref{thm: integer fact} has, as Step 3, a
check for repetitions amongst the set $S$ only at the initial construction of the
graph.  However, this point could be exploited further at future diffusion steps, and
the detection of such a repetition would detect the order $r$.  For example, if
the order of the element is $561$, then by writing $561=512+32+16+1$, we will
have a repetition on the fourth diffusion step.  Also, if the order of the element is $111$, then since $111 = 128 - 16 -1$,
one would have a repetition on the second diffusion
step.  In general, there will be a repetition after $n+1$ steps if and only if
one has $a^{k} \equiv a^{\ell} \textrm{\rm \, mod \,}N$ for distinct integers $k$ and $\ell$
each of whose binary expansion consists of $h$ or fewer non-zero digits.

\vskip .10in
In Step 3 of Theorem \ref{thm: integer fact}, we used $O(\log N)$ digital steps
to detect repetitions. This investigation occurred at the first diffusion step.
The naive method to test for further repetitions at
diffusion step $h$ would require $O((\log N)^{h})$ digital steps.  At this time, we have
not devised a diffusion algorithm which would more efficiently undertake the problem
of seeking higher repetitions.

\subsection{Diffusion solutions of the Simon and Deutsch-Jozsa problems}

For each integer $h \geq 2$, the Hamming cube, or hypercube graph, $Q_{h}$ of dimension $h$
is constructed as follows.  The vertices correspond to $h$-tuples where each entry
is either $0$ or $1$.  There are $2^{h}$ vertices.  An edge is formed by connecting
two vertices $v_{1}$ and $v_{2}$ if and only if $v_{1}$ and $v_{2}$ differ in one and
only one place.  There are $2^{h-1}h$ edges.

\vskip .10in
In \cite{HoRe20} the discrete time heat kernel $p_{n}^{(h)}(v)$ on $Q_{h}$ is studied,
and its spectral expansion is explicitly stated.  With this information,
Simon's problem and the Deutsch-Jozsa problem are studied in Chapter 4 of \cite{HoRe20}.
To recall, the statement of these problems are as follows.

\vskip .10in
\begin{enumerate}
\item {\bf The Deutsch-Jozsa problem.} Let $f$ be a Boolean function on the vertices
of $Q_{h}$,
meaning the range of values of $f$ is $\{0,1\}$.  Suppose we know that $f$ is either
constant or balanced, by which we mean that the pre-image of either value $0$ or
$1$ is one-half of the vertices of $Q_{h}$.  Determine the number of steps required to
decide if $f$ is constant or balanced.
\item {\bf Simon's problem.} Let $f$ be a real-valued function on the vertices of $Q_{h}$.
Let us view $v_{1}$ and $v_{2}$ as vectors of integers $\textrm{\rm mod \rm} 2$.  Assume
there is a vertex $s$, meaning an $h$-tuple $\textrm{\rm mod \,}2$, such that $f(v_{1})=f(v_{2})$ if and only
$v_{1}+v_{2} \equiv 0 \textrm{\rm \,mod\,} 2 $ or $v_{1}+v_{2} \equiv 0 \textrm{\rm \,mod \,} 2 $.  Determine the number of steps required to compute $s$.
\end{enumerate}

\vskip .10in
Using a digital computer, the Deutsch-Jozsa problem requires, in the worst case, $2^{h-1}+1$ evaluations  to be
solved.  On a quantum computer, the solution is obtained after a single quantum step.  It is
shown in section 4.4.1 of \cite{HoRe20} that a \it heat computer \rm constructed from discrete time heat diffusion on $Q_{h}$ provides a heat computer solution to the Deutsch-Jozsa problem in a single heat step, which means the same as a diffusion step.

\vskip .10in
Using a probabilistic approach, Simon's problem can be solved on a digital computer in
$O(2^{h/2})$ steps, whereas on a quantum computer a solution is obtained in $O(h)$ quantum
steps.  Similarly, it is shown in section 4.4.2 of \cite{HoRe20} that a heat computer
can solve Simon's problem in $O(h)$ diffusion steps.

\vskip .10in
We find it fascinating that a diffusion computer, or heat computer as the concept
was called in \cite{HoRe20}, can efficiently solve the Deutsch-Jozsa problem and Simon's problem.
In addition, the number of diffusion steps in the solutions coincide with the number of
quantum steps needed to answer the questions using a quantum computer.  In that regard,
Theorem \ref{thm: integer fact} can be compared to Shor's algorithm.

\vspace{5mm}\noindent
Carlos A. Cadavid \\
Department of Mathematics \\
Universidad Eafit \\
Carrera 49 No 7 Sur-50 \\
Medell\'in, Colombia \\
e-mail: ccadavid@eafit.edu.co

\vspace{5mm}\noindent
Paulina Hoyos  \\
Department of Mathematics \\
The University of Texas at Austin \\
C2515 Speedway, PMA 8.100 \\
Austin, TX 78712
U.S.A. \\
e-mail: paulinah@utexas.edu

\vspace{5mm}\noindent
Jay Jorgenson \\
Department of Mathematics \\
The City College of New York \\
Convent Avenue at 138th Street \\
New York, NY 10031
U.S.A. \\
e-mail: jjorgenson@mindspring.com

\vspace{5mm}\noindent
Lejla Smajlovi\'c \\
Department of Mathematics \\
University of Sarajevo\\
Zmaja od Bosne 35, 71 000 Sarajevo\\
Bosnia and Herzegovina\\
e-mail: lejlas@pmf.unsa.ba

\vspace{5mm}\noindent
Juan D. V\'elez \\
Department of Mathematics \\
Universidad Nacional de Colombia\\
Carrera 65 Nro. 59A - 110\\
Medell\'in, Colombia\\
e-mail: jdvelez@unal.edu.co 
  
\newpage  
\section*{Appendix: Computer code in \it Maple \rm}

\vskip .10in
For the reader convenience we present the \it Maple \rm implementation of the algorithm described in Section \ref{sec: Proof of Thm integer fact}, which we used to do the examples in Section  \ref{Ejemplos}. For a better understanding we show in detail the steps for carrying out Example 2. \\

with(numtheory):\\
with(ArrayTools): \\
with(Statistics): \\

ArrayAsList := proc(ArrayWithTwoColumns) local A, s, List, i; A := ArrayWithTwoColumns; s := Size(A, 1); List := [ ]; for i to s do List := [op(List), [A[i, 1], A[i, 2]]]; end do; return List; end proc \\

Joint := proc(ArrayWithTwoColumns) local M, i, z, C, A, s, T, P, NA; A := ArrayWithTwoColumns; s := Size(A, 1); T := ArrayAsList(A); M := sort(T, (x, y) $\rightarrow$ x[1] $\leq$ y[1]); P := convert(M, Array); NA := Array(1 .. 1, 1 .. 2); NA[1, 1] := P[1, 1]; NA[1, 2] := P[1, 2]; for i from 2 to s do z := Size(NA, 1); if M[i][1] = NA[z, 1] then NA[z, 2] := M[i][2] + NA[z, 2]; else NA := Concatenate(1, NA, Vector[row]([M[i][1], M[i][2]])); end if; end do; return NA; end proc \\

FromMultisetToList := proc(Multiset) local List, M, T, i, j; List := []; M := Entries(Multiset); T := nops(M); for i to T do for j to M[i][2] do List := [op(List), M[i][1]]; end do; end do; return List; end proc: \\

N := 47*29 \\

a := 991 \\

M := floor(log[2](N)) + 1 \\

         M:=11\\

FormS := proc( c ) local i, SeparatedList; SeparatedList := [[c], [1/c mod N]];
for i to M do SeparatedList := [[op(SeparatedList[1]), (SeparatedList[1][i])$^2$  mod N],
[op(SeparatedList[2]), SeparatedList[2][i]$^2$ mod N]]; end do; return SeparatedList; end proc \\

Sa := FormS(a)

Sa := [[991, 721, 538, 488, 982, 683, 343, 431, 393, 430, 895, 944], [905, 1225, 1325, 81, 1109, 455, 1212, 993, 600, 168, 964, 1093]]

ConectS := proc(S) local UniqueList, j; UniqueList := S[1]; for j to M + 1 do UniqueList := [op(UniqueList), S[2][j]]; end do; return UniqueList; end proc \\

Ta := ConectS(Sa)\\

Ta := [991, 721, 538, 488, 982, 683, 343, 431, 393, 430, 895, 944, 905, 1225, 1325, 81, 1109, 455, 1212, 993, 600, 168, 964, 1093]\\

FirstRepetition := proc(List) local long, H, i, j, K; K := []; long := nops(List); H := 0; for i to long do for j to i - 1 while H = 0 do if List[i] = List[j] then K := [j, i]; H := 1; end if; end do; end do; return K; end proc \\

FirstRepetitionInTwoListsOfSameLength := proc(L, M) local Repetition, u, i, Rep1, j, d1, Rep2, k, d2; Repetition := []; u := nops(L); for i to u while Repetition = [] do Rep1 := []; for j to i while Rep1 = [] do if L[i] = M[j] then Rep1 := [i, j]; d1 := j + i; Repetition := [Rep1]; end if; end do; Rep2 := []; for k to i while Rep2 = [] do if M[i] = L[k] then Rep2 := [k, i]; d2 := k + i; Repetition := [op(Repetition), Rep2]; end if; end do; end do; if nops(Repetition) = 0 then return []; end if; if nops(Repetition) = 2 then if d1 <= d2 then Repetition := [Repetition[1]]; else Repetition := [Repetition[2]]; end if; end if; return op(Repetition); end proc \\

RepetitionAndFactors := proc(S) local PositPot, NegAndPositPot, Answer, d1, d2, l1, l2, k1, k2, r, q, pot, u, v, t, h, b, f1, f2; PositPot := FirstRepetition(S[1]); NegAndPositPot := FirstRepetitionInTwoListsOfSameLength(S[1], S[2]); if PositPot = [ ] and NegAndPositPot = [ ] then return No repetition; else if PositPot = [ ] and NegAndPositPot $\neq$ [ ] then u := NegAndPositPot[1] - 1; v := NegAndPositPot[2] - 1; r := min(u, v); t := max(u, v); q :=$2^{t-r}+1$; else if PositPot $\neq$ [ ] and NegAndPositPot = [ ] then u := PositPot[1] - 1; v := PositPot[2] - 1; r := min(u, v); t := max(u, v); q := $2^{t-r}-1$; else d1 := abs(PositPot[1] - PositPot[2]); d2 := abs(NegAndPositPot[1] - NegAndPositPot[2]); if d1 $\leq$ d2 then Answer := [op(PositPot[1]), op(PositPot[2])]; else Answer := [op(NegAndPositPot[1]), -op(NegAndPositPot[2])]; if Answer[2] $<$ 0 then l1 := abs(Answer[1]) - 1; l2 := abs(Answer[2]) - 1; k1 := min(l1, l2); k2 := max(l1, l2); r := k1; q := $2^{k2 - k1} + 1$; else l1 := Answer[2] - 1; k1 := min(l1, l2); k2 := max(l1, l2); r := k1; q := $2^{k2 - k1} - 1$; end if; end if; end if; end if; end if; pot := r; for h from 0 to r while $a^{2^{pot}q}$ = 1 mod N do pot := pot - 1; end do; b := $2^{(pot - 1)q}$; f1 := gcd(N, $a^b$ + 1); f2 := gcd(N, $a^b$ - 1); return f1, f2; end proc \\

RepetitionAndFactors(Sa)\\

No repetition\\

b:= $a^{2^M}$ mod N\\

b:= 944 \\

Tb := ConectS(FormS(b)) \\

Tb:=[944, 1097, 1243, 770, 1358, 25, 625, 807, 1098, 712, 1271, 286, 1093, 661, 761, 1209, 545, 1254, 977, 429, 36, 1296, 400, 529] \\

InitialNumberOfHeatUnits := 1 \\

OneDiffusionStep := proc(ArrayWithTwoColumns) local NewMultiset, R, Q, i, j, H, U, A, NA; A := ArrayWithTwoColumns; NA := Array(1 .. 1, 1 .. 2); R := Size(A, 1); Q := 2*M + 2; for i to R do NA := Concatenate(1, NA, Vector[row]([A[i, 1], 1/2*A[i, 2]])); for j to Q do NA := Concatenate(1, NA, Vector[row]([A[i, 1]*Tb[j] mod N, 1/2*A[i, 2]/Q])); end do; end do; U := Joint(NA); return U[2 .. ]; end proc \\

SeveralDiffusionSteps := proc(k) local P, i, Z; P := Array(1 .. 1, 1 .. 2, [[1, InitialNumberOfHeatUnits]]); for i to k do P := OneDiffusionStep(P); end do; return P; end proc \\

distrib := SeveralDiffusionSteps(25) \\

\begin{figure}[h!]
	\centering
	\includegraphics[scale=1]{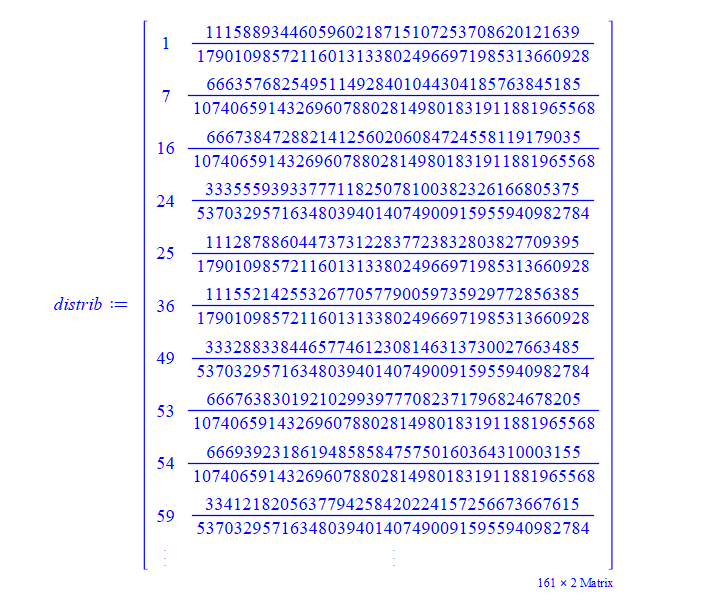}
\end{figure}

ListToGraph := ArrayAsList(distrib)\\

TransfReciprocals := proc(Lista) local NuevaLista, i; NuevaLista := []; for i to nops(Lista) do NuevaLista := [op(NuevaLista), evalf(1/Lista[i][2])]; end do; return NuevaLista; end proc \\

RR := TransfReciprocals(ListToGraph) \\

Histogram(RR, bincount = 11) 

\begin{figure}[h!]
	\centering
	\includegraphics[scale=0.6]{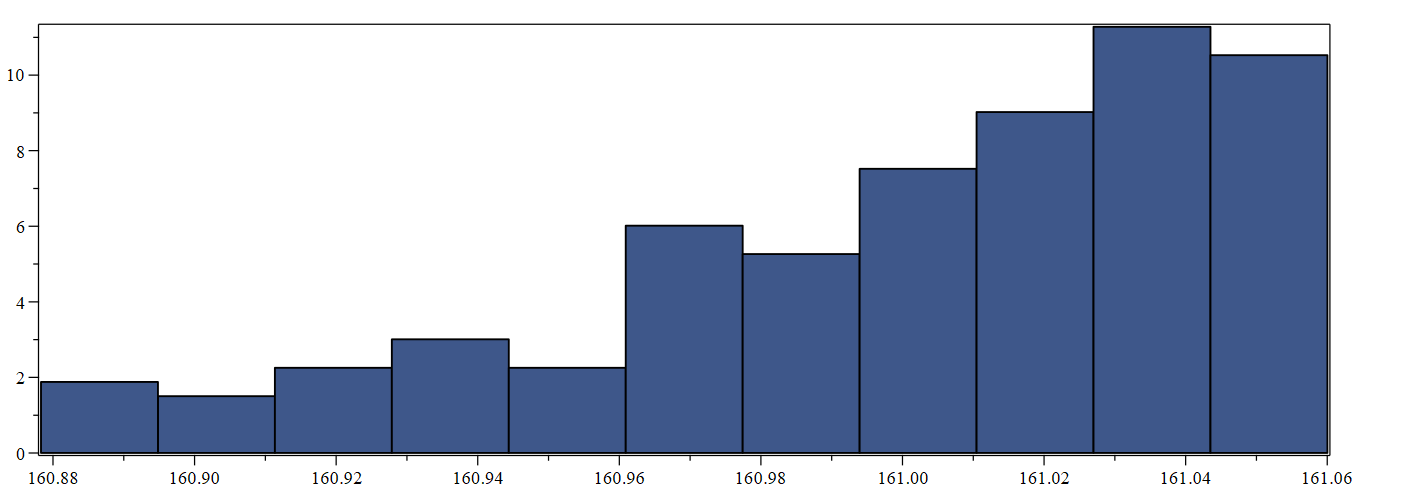}
 \caption{Histogram for the reciprocals of probabilities after 36 diffusion steps}
\end{figure}

\end{document}